\documentstyle[11pt,aaspp4,nick]{article}
\begin{document}


\title{Formation of Galactic Bulges}
\author{Nickolay Y.\ Gnedin\altaffilmark{1},
Michael L.\ Norman\altaffilmark{2}, and
Jeremiah P.\ Ostriker\altaffilmark{3}}
\altaffiltext{1}{Center for Astrophysics and Space Astronomy, 
University of Colorado, Boulder, CO 80309;
e-mail: \sl gnedin@casa.colorado.edu}
\altaffiltext{2}{Laboratory for Computational Astrophysics, 
National Center for Supercomputing Applications, 
University of Illinois at Urbana-Champaign, 
405 North Matthews Avenue, Urbana, IL 61801; 
e-mail: \sl norman@ncsa.uiuc.edu}
\altaffiltext{3}{Princeton University Observatory, Peyton Hall, 
Princeton, NJ 08544; e-mail: \it jpo@astro.princeton.edu} 


\load{\scriptsize}{\sc}

\def\A{{\cal A}}
\def\B{{\cal B}}
\def\ion#1#2{\rm #1\,\sc #2}
\def\HI{{\ion{H}{i}}}
\def\HII{{\ion{H}{ii}}}
\def\GI{{\ion{He}{i}}}
\def\GII{{\ion{He}{ii}}}
\def\GIII{{\ion{He}{iii}}}
\def\MH{{{\rm H}_2}}
\def\Hp{{{\rm H}_2^+}}
\def\Hm{{{\rm H}^-}}

\def\dim#1{\mbox{\,#1}}

\def\figdir{.}
\def\placefig#1{#1}

\begin{abstract}
We use cosmological hydrodynamic simulations to investigate formation
of galactic bulges within the framework of hierarchical clustering
in a representative CDM cosmological model. We show that largest objects
forming at cosmological redshifts $z\sim4$ resemble observed bulges
of spiral galaxies or moderate size ellipticals
in their general properties like sizes, shapes,
and density profiles. This is consistent with observational data indicating the
existence of ``old'' bulges and ellipticals at more moderate redshifts.
These bulges are gas dominated at redshift $z=3$, with high rates of star formation and would appear to be good candidates for small blue galaxies seen
in the Hubble Deep Field.
\end{abstract}

\keywords{cosmology: theory - cosmology: large-scale structure of universe -
galaxies: formation - galaxies: intergalactic medium}

\section{Introduction}

The fact that bulges of ordinary galaxies (and indistinguishable ellipticals 
of the same luminosity) are very dense is often used
as an argument against the currently favored CDM-type cosmological
models. Really, since the average density of the universe decreases
with time, and the average density of a bound object is directly proportional
to the density of the universe at the time when the object is formed,
dense galactic bulges should have formed at very high redshift.
Thus, it is reasonable to ask whether 
currently fashionable cosmological models normally have enough
small scale power to account for the formation of massive 
($10^9-10^{10}$ solar masses in baryons) bulges at $z\ga10$ (Peebles 1997).

This argument can be illustrated by the following simple estimate: 
the characteristic
number density of baryons in the Galactic bulge within the sphere with
the radius of $3\dim{kpc}$ is
\begin{equation}
	n_{\rm GB} \sim 2 \dim{cm}^{-3}
	\label{ngb}
\end{equation}
(Kent, Dame, \& Fazio 1991; Freudenreich 1998).
In comparison, the average density of the cosmological virialized object
formed at redshift $z$ is only
\begin{equation}
	n_{\rm TH} \sim 10^{-4} (1+z)^3 \dim{cm}^{-3}
	\label{nth}
\end{equation}
for $\Omega_bh^2=0.02$, and we assume that the average density of the 
virialized
object is about 200 times the average density of the universe at the moment of
formation (Gunn \& Gott 1972), according to the 
standard dissipationless collapse theory.
Comparing equations (\ref{ngb}) and (\ref{nth})
we can deduce that the bulge of our Galaxy formed at $z_{\rm GB}
\sim 30$. But, none of the currently acceptable CDM-type models can form
 $10^{10}$ solar mass baryonic objects at $z=30$ in numbers even
closely
comparable to the observed number density of galactic bulges.

Does this argument imply that the CDM-type models are ruled out? We will
try to show in this paper that the answer to this question is no. What
this simple argument misses is the ability of baryons to cool and
collapse to the densities exceeding that of the dark matter. We present
a series of cosmological hydrodynamic simulations which include
adequate physical modelling to properly account for formation of
bulges of galaxies, and we show that typical bulges form in a
realistic CDM-type model at $z\sim5$ rather than at $z\sim30$.

But before describing our detailed results, it is perhaps worth noting the
conceptual flaw in the simple argument we first presented. The global ratio
of baryons to dark matter in our simulation, and in typical current estimates
is about 1:8. But in our own galactic bulge the baryonic -- stellar --
component exceeds the dark matter component and may exceed it by as much
as a factor of a few. Thus, (taking the cube root of density enhancement)
the baryonic stellar component has contracted relative
to the dark matter component by a factor of two to four and, correspondingly,
the bulge should have formed in the plausible redshift range of
$z\sim 5-10$ rather than at $z=30$. Furthermore, as we shall see, the
observed (and computed) profiles are very different from those envisioned
in the top-hat collapse picture.

An aside on a related but quite different problem may be useful here. There 
is currently a good deal of discussion on the subject of whether or not 
standard CDM models make bulges which are {\it too\/} dense (e.g.\ Spergel
\& Steinhardt 1999, Burkert \& Silk 1999, Moore et al.\ 1999, 
Kravtsov et al.\ 1998). However, these papers
address far lower mass systems and ones which are dark matter dominated,
having circular velocities of less than $100\dim{km/s}$. 
Here we address normal bulges
with (equivalent) circular velocities of about $200\dim{km}/\dim{s}$ or
more, for which the advertised problem has been the opposite: why are
they {\it so\/} dense?

Some work on the subject was done by Steinmetz \& 
Mueller (1994, 1995). However,
they modeled the formation of disk galaxies as a collapse of an isolated gas 
cloud, whereas in this work we consider the formation of galactic bulges 
within the framework of hierarchical clustering, based on the realistic 
cosmological simulations that include such effects as the cosmological infall, 
merging of proto-galactic clumps, and expansion of the universe, all of which 
are missing in the Steinmetz \& Mueller work. Other numerous numerical 
investigations of galaxy formation do not specifically concentrate on the
bulges, or lack mass and/or spatial resolution to say anything about
bulge formation.

\section{Simulations}

We use the SLH cosmological hydrodynamic code (Gnedin 1995, 1996; Gnedin \&
Bertschinger 1996). Physical modelling included in the code is fully
described in  Gnedin \& Ostriker (1997). We choose a CDM+$\Lambda$
cosmological model with the following cosmological parameters:
$$
	\Omega_0 = 0.37,\ \ \ \Omega_L = 0.63,\ \ \ h = 0.70,\ \ \ 
	\Omega_b = 0.049,
$$
which is close to the ``concordance'' model of Ostriker \& 
Steinhardt (1995) and to the models consistent with recent SNIa results
(Riess et al.\ 1998, Perlmutter et al.\ 1999).
We have performed one simulation with $128^3$ baryonic resolution elements,
the same number of dark matter particles, and a number of stellar particles
were formed during the simulation. The simulation box size was fixed to
$3h^{-1}{\rm\,Mpc}$, which resulted in the total mass resolution of
$1.3\times10^6h^{-1}\dim{M}_{\sun}$. For reference, the Jeans mass at
$z=10$ and $T=10^4\dim{K}$ is about $10^{10}h^{-1}\dim{M}_{\sun}$.
The spatial resolution was fixed at
$1.5h^{-1}$ comoving kiloparsecs. 
Because of the small box size, this simulation
cannot be continued to $z=0$. Instead, we stopped the simulation at 
$z\approx3$. 

\def\capCD{Evolution of the central dark matter density ({\it solid lines\/}),
baryonic density ({\it dashed lines\/}), and stellar density
({\it dotted lines\/}) for the most massive object in our simulation. 
The thin solid line shows the central total
density of the same object if it had the Navarro-Frenk-White (NFW)
density profile. The central core appears because of finite spatial
resolution in simulations.}
\placefig{
\begin{figure}
\insertfigure{\figdir/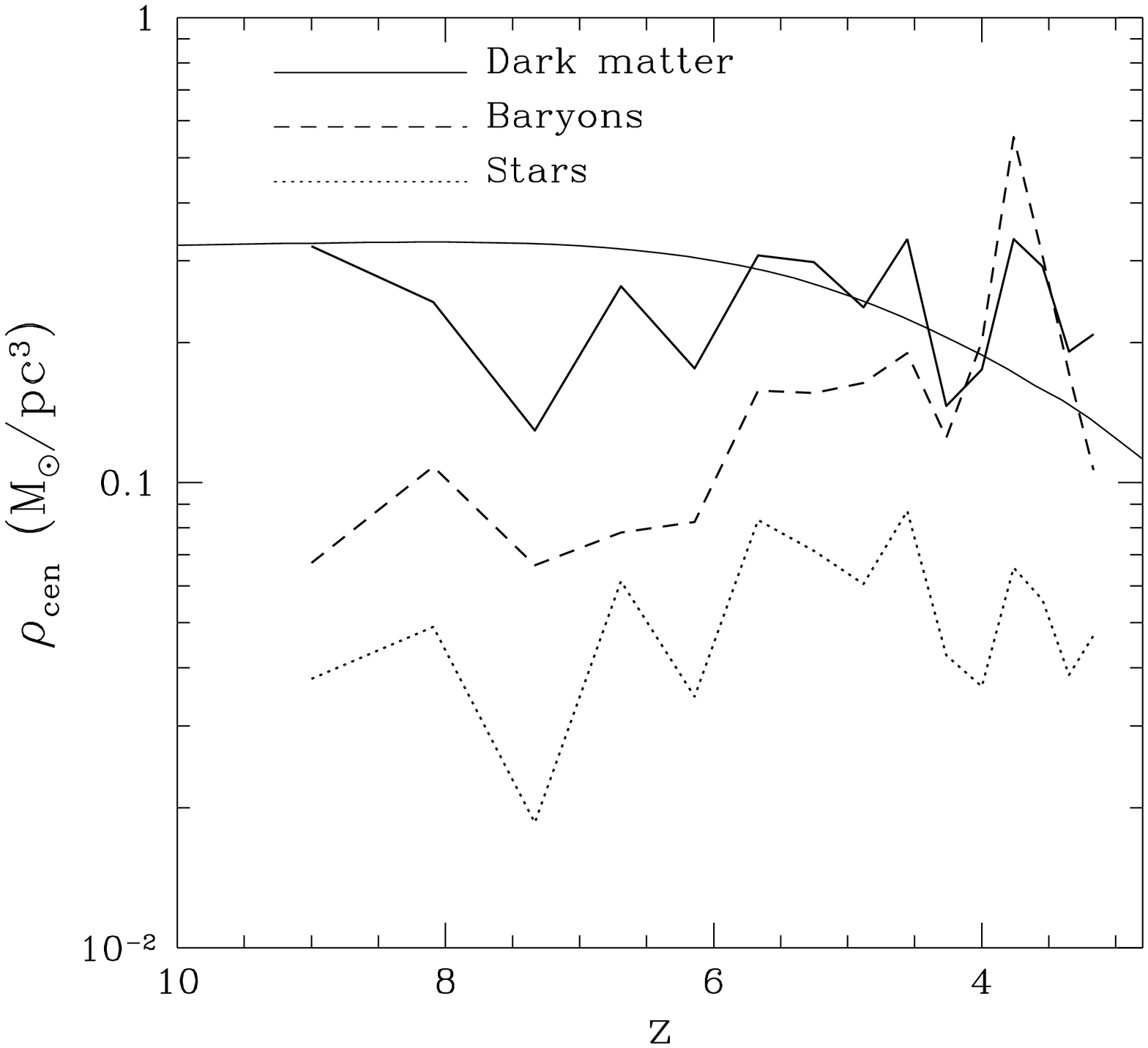}
\caption{\label{figCD}\capCD}
\end{figure}
}
Even at this redshift our simulation is suffering from
the lack of spatial of resolution, as can be illustrated by Figure \ref{figCD},
which shows the evolution of the central density for the dark
matter, baryonic and stellar components of the most massive object from
our simulation. The central density
is defined as the average density within the sphere of two resolution
lengths of our simulations ($\sim 900\dim{pc}$ at $z=3.8$). As an 
illustration, we also show the central density of the Navarro, Frenk, and
White (NFW) profile, which is defined in the same way and is therefore a 
function of
resolution (and thus time, as we keep the resolution fixed in the comoving
coordinates) - as redshift decreases, so does our spatial resolution in the
physical coordinates. The fact that the NFW density actually starts to decrease
for $z<6$ indicates that our resolution becomes comparable to the 
characteristic radius of the object - the radius where the local density slope
approaches -2.

Another simulation with eight times more resolution elements ($256^3$) and
$6h^{-1}{\rm\,Mpc}$ box was also performed. The large simulation thus
had the same mass resolution as the small one, and the spatial resolution
in the large simulation was fixed at $1.2h^{-1}$ comoving kiloparsecs.
However, because the large simulation required a computational expense
beyond what was available to us, it was terminated at $z=9.5$. Thus,
we used the large simulation to verify numerical convergence and
estimate missing small scale power, but we will use the small ($128^3$)
simulation as the source for scientific results. 

By comparing the large and small simulation, we have found that the small
simulation included most of the small scale power that was initially present
in the baryonic component. Thus, our results are not significantly
affected by the finite resolution in the initial conditions
($k$-space resolution), but they are, of course, subject to finite
mass and spatial resolution.

\section{Results}

Since we are concerned with the process of formation of galactic
bulges and small ellipticals, we will focus in this paper on properties of individual
objects formed in our simulations. Specifically, we will focus
on four most massive objects. Each of those objects contains more
than ten thousand particles of each kind (i.e.\ the dark matter,
gas, and stars), and thus they are fully resolved numerically.

\def\tableone{
\begin{table}
\caption{Four Most Massive Objects at $z=4$}
\medskip
$$
\begin{tabular}{cccccc}
\label{tabone}
Object & Total mass ($\dim{M}_{\sun}$) & Baryonic mass ($\dim{M}_{\sun}$) & 
Stellar mass ($\dim{M}_{\sun}$) & $M_b/M_t$ & $M_*/M_b$ \\ \tableline
 A & $4.0\times10^{10}$ & $8.2\times10^{9}$ & $1.5\times10^{9}$&0.20&0.18 \\ 
 B & $2.9\times10^{10}$ & $5.1\times10^{9}$ & $1.2\times10^{9}$&0.18&0.23 \\ 
 C & $1.8\times10^{10}$ & $3.3\times10^{9}$ & $0.9\times10^{9}$&0.18&0.26 \\ 
 D & $1.4\times10^{10}$ & $2.2\times10^{9}$ & $0.6\times10^{9}$&0.16&0.29 \\ 
\end{tabular}
$$
\end{table}
}
\def\tableoneb{
\begin{table}
\caption{Four Most Massive Objects at $z=3$}
\medskip
$$
\begin{tabular}{cccccc}
\label{taboneb}
Object & Total mass ($\dim{M}_{\sun}$) & Baryonic mass ($\dim{M}_{\sun}$) & 
Stellar mass ($\dim{M}_{\sun}$) & $M_b/M_t$ & $M_*/M_b$ \\ \tableline
 A & $7.1\times10^{10}$ & $1.1\times10^{10}$ & $2.4\times10^{9}$&0.15&0.23 \\ 
 B & $2.9\times10^{10}$ & $4.5\times10^{9}$  & $1.2\times10^{9}$&0.16&0.27 \\ 
 C & $1.8\times10^{10}$ & $3.0\times10^{9}$  & $0.9\times10^{9}$&0.17&0.22 \\ 
 D & $1.6\times10^{10}$ & $2.5\times10^{9}$  & $0.9\times10^{9}$&0.16&0.34 \\ 
\end{tabular}
$$
\end{table}
}
\placefig{\tableone}
\placefig{\tableoneb}
Tables \ref{tabone} and \ref{taboneb} present the general properties of the 
four objects: their
total, baryonic, and stellar masses, as well as the ratio of the baryonic
to the total mass, and the stellar to the baryonic mass at $z\approx4$ and 
$z\approx3$
respectively.

A few observations can easily be made from the table. First, the
objects contain more baryons than the cosmic average of 
$\Omega_b/\Omega_0=0.13$. Second, they are still dominated by gas, 
as only about
20--30\% of their baryonic mass is turned into stars when we terminated
the computation at $z\approx3$. These objects have a moderately high rate of 
star formation (several solar masses per year), and would appear to correspond
well to the numerous small blue objects seen in the Hubble Deep Field
(Contardo, Steinmetz, \&  Alvensleben 1998).

But, it must be noted here that the amount
of gas turned into stars really depends on the specifics of the star
formation algorithm adopted in the simulation. We make no pretense that
we can fully account for the star formation in our simulations, it is
clear that modeling of star formation needs to be much advanced before
simulations could claim to predict the star formation rate in 
(proto-)galaxies with sufficient detail. Thus, for the purpose of this
paper we will pay little attention to the stellar component of our
simulated bulges, and will focus primarily on the total baryonic component
instead. We assume that with proper star formation algorithm, our
simulated bulges will form stars at an appropriate rate as long as
the total baryonic distribution is compatible with observations.

\def\capDP{Density profiles for the dark matter ({\it solid lines\/}),
total baryons (gas and stars, {\it dashed lines\/}), and stars
({\it dotted lines\/}) of the 
four most massive objects at $z=3.8$ as a function
of radius in physical (not comoving) units. The bold 
vertical bar marks the spatial resolution of the simulation (450 pc).}
\placefig{
\begin{figure}
\insertfigure{\figdir/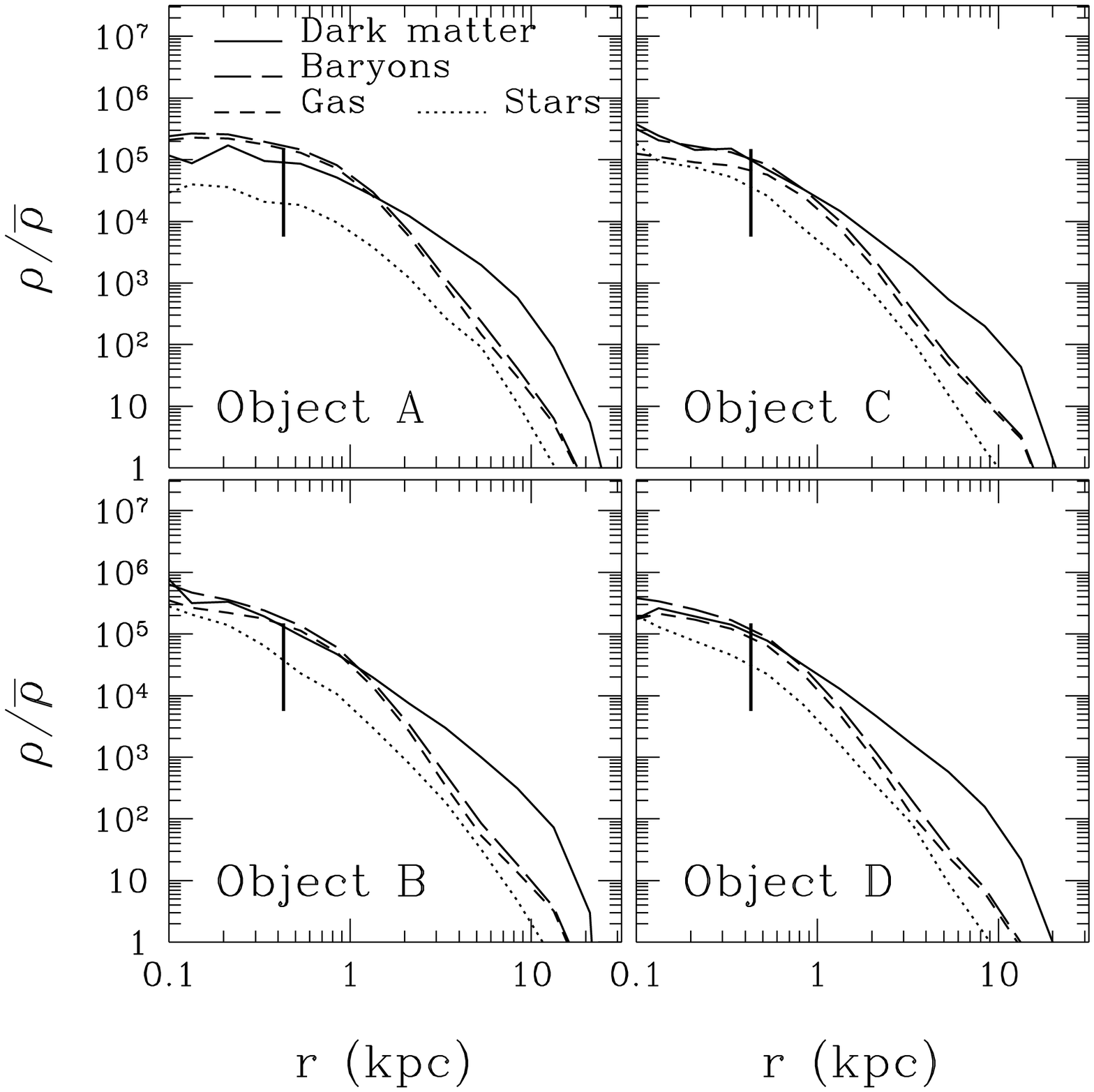}
\caption{\label{figDP}\capDP}
\end{figure}
}
Since at $z<3.8$ our results become substantially contaminated
by the lack of spatial resolution, we will restrict most of our
analysis to the redshift $z=3.8$, the lowest redshift at which
characteristic radii of bound objects are still well resolved.
Density profiles of the four most massive objects are shown in Figure
\ref{figDP}. The most massive object, object A, is less dense at the center
than other three objects because it experienced a major merger shortly
before $z=4$ and has not fully relaxed yet. Object D has also experienced
a major merger at $z\approx6$, whereas objects B and C have been accreting
matter quietly since $z\sim 10$. We point out here that in all four objects
baryonic density at the center is greater than or similar to the dark matter 
density,
i.e.\ baryons in all four objects are 
self-gravitating. As an aside we note that should the gaseous component 
dominate at the center at any time, a rapid collapse would occur which
presumably would be accompanied by rapid star formation or, should a black 
hole of sufficient mass be presents, the concomitant flare up of an AGN.
Clearly, we have not sufficient resolution to follow this phase
(but see Abel, Bryan, \& Norman 1998).

\def\capMZ{Evolution of the total mass ({\it solid lines\/}),
baryonic fraction ({\it dashed lines\/}), and stellar fraction
(per unit baryonic mass, {\it dotted lines\/}) of the
four most massive objects.}
\placefig{
\begin{figure}
\insertfigure{\figdir/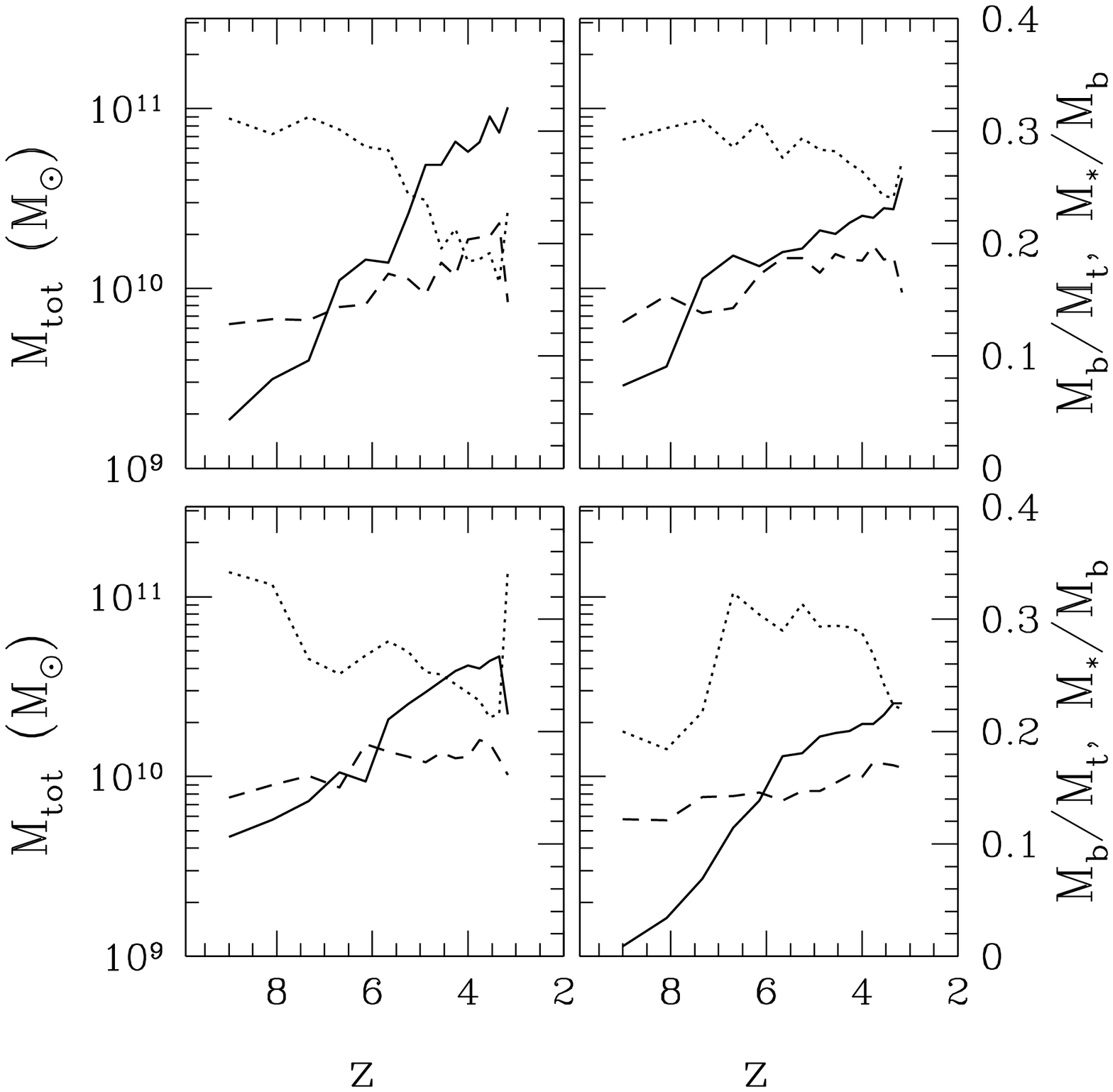}
\caption{\label{figMZ}\capMZ}
\end{figure}
}
Evolution of average properties of these objects is shown in Figure
\ref{figMZ}. One can immediately see that all four objects are experiencing
heavy merging at $z\sim4-6$, increasing their mass by about an order of
magnitude. The stellar fraction of object A decreased significantly at
$z=4-5$ because of accretion of a large quantity of fresh gas. Also
noticeable is the increase in the total baryonic fraction in object A
at $z\sim4$. Since baryons are almost self-gravitating at the center of
object A, their efficiently cool and collapse toward the center, leading
to the increase in the baryonic fraction of the object.

Figure \ref{figCD} shows the evolution of the central density for the dark
matter, baryonic and stellar components of object A. One can see
that the dark matter density does not change systematically with time
(albeit fluctuating significantly) until the lack of resolution contaminates
results, because the object is close to
the virial equilibrium. On the contrary, the baryonic density increases
with time because of efficient cooling at the center of the object. 
The recent merger at $z\sim5$ triggered a considerable increase in the
central density of gas, but this increase has not yet resulted in the
burst of star formation. We expect, that if we continued the simulation
to lower redshift with higher spatial resolution, 
object A would experience a burst of star formation at
the center, which would transform most of the gas into stars on a rather
short time-scale. 

\def\capSP{Surface mass density profiles for the dark matter 
({\it solid lines\/}) and total baryons ({\it dashed lines\/}) 
of the four most massive objects at $z=3.8$ as a 
function of radius in physical (not comoving) units. The profiles are
terminated at the resolution limit of the simulation (450 pc). Tilted 
long-dashed lines show $r^{1/4}$ law for the baryonic profiles, and 
dotted lines show exponential profiles. Bold arrows show characteristic
radii $R_e$ for $r^{1/4}$ profiles.
The 
horizontal long-dashed line show the central surface density of a homogeneous
top-hat sphere.}
\placefig{
\begin{figure}
\epsscale{0.70}
\insertfigure{\figdir/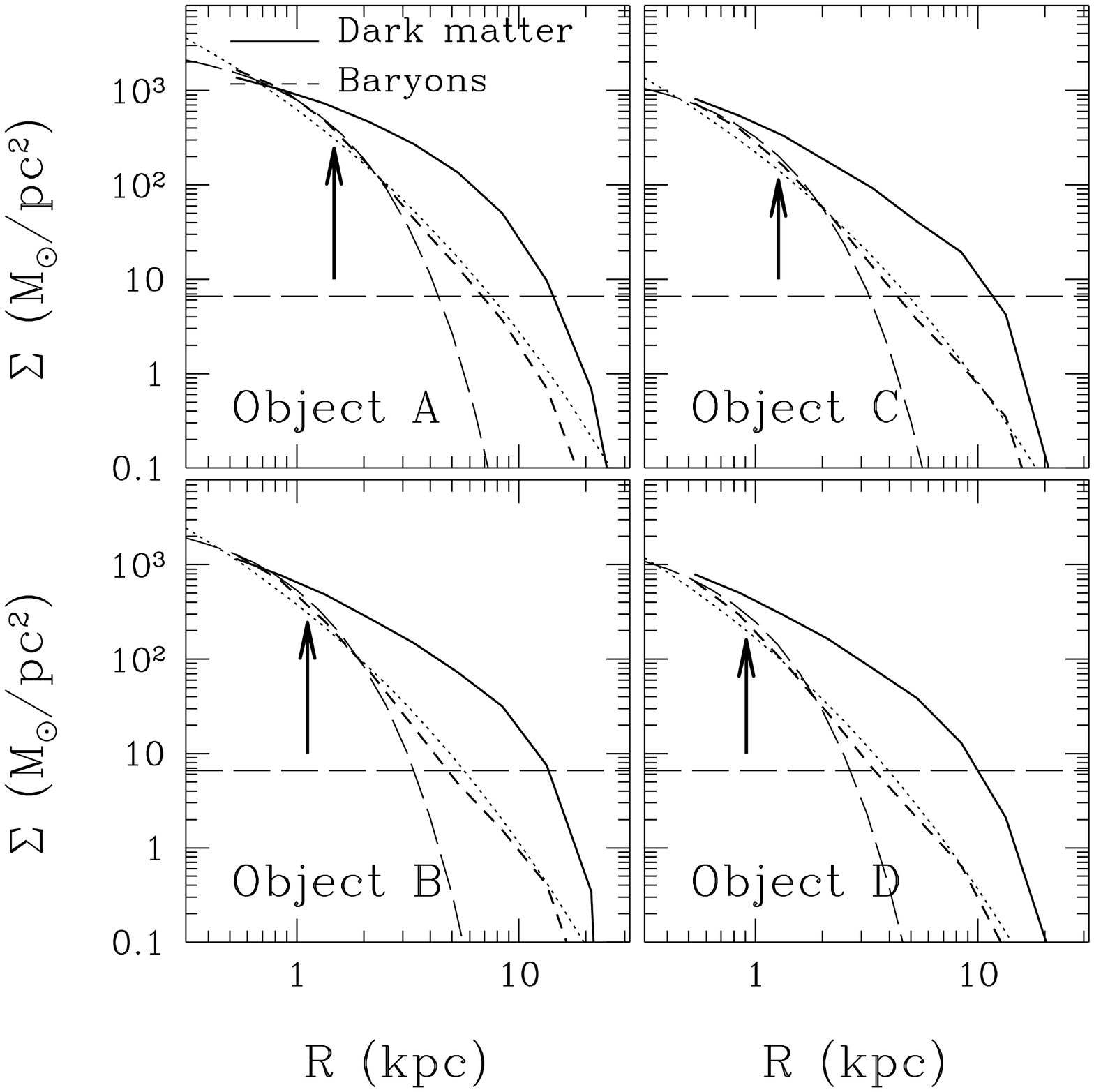}
\caption{\label{figSP}\capSP}
\end{figure}
}
We are now ready to address the major question of this paper: are those
objects formed in the simulation resemble real galactic bulges? In order
to answer this question, we show in Fig.\ \ref{figSP} the surface density
profiles for our four objects. Also, for the total baryonic profile
we compute the exponential fits in the form:
\begin{equation}
        \Sigma(R) = \Sigma_C e^{\displaystyle 1-R/R_C}
	\label{expfit}
\end{equation}
and $r^{1/4}$ law fits,
\begin{equation}
        \Sigma(R) = \Sigma_e 10^{-3.3307\left((R/R_e)^{1/4}-1\right)}.
	\label{rqlfit}
\end{equation}
As can be seen from Fig.\ \ref{figSP} the central $3\dim{kpc}$ of the
simulated bulges are equally well fitted both by the exponential and
by the de Vaucouleurs profiles, and we lack the resolution to distinguish
them at the very center. We also note that our resolution is barely
enough to resolve the characteristic scales of the two profiles, so
we may overestimate the characteristic radii somewhat, but not
by a large factor. However, 
both, Kent, Dame \& Fazio (1991) and Freudenreich (1998) models for the
galactic bulge are well fitted by the exponential profile for the range
$0.1\dim{kpc}<r<3\dim{kpc}$ (with respective rms errors of 3 and 8 percent
respectively), and thus agree with our simulations over the range of scales
which we can resolve.

\def\tabletwo{
\begin{table}
\caption{Fit Parameters for the Four Most Massive Objects at $z=3.8$}
\medskip
\begin{tabular}{ccccc}
\label{tabtwo}
Object & $R_C$ (kpc) & $\Sigma_C$ ($\dim{M}_{\odot}/\dim{pc}^2$) & $R_e$ (kpc) & $\Sigma_e$ ($\dim{M}_{\odot}/\dim{pc}^2$) \\ \tableline
 A             & 0.70 & 1200 & 1.47 & 310 \\
 B             & 0.54 & 1270 & 1.11 & 310 \\
 C             & 0.58 & 664  & 1.27 & 140 \\
 D             & 0.46 & 790  & 0.91 & 200 \\
Galactic bulge\tablenotemark{a} & 0.58 & 650\tablenotemark{c} & 0.97 & 250\tablenotemark{c} \\
Galactic bulge\tablenotemark{b} & 0.47 & 560\tablenotemark{c} & 0.76 & 260\tablenotemark{c} \\
\tablenotetext{a}{Kent, Dame, \& Fazio 1991}
\tablenotetext{b}{Freudenreich 1998}
\tablenotetext{c}{A mass to light ratio of 3 is assumed in converting 
luminosity to mass.}
\end{tabular}
\end{table}
}
\placefig{\tabletwo}
We put together the parameters of the fits in Table \ref{tabtwo}. In addition,
we list the parameters that describe the Galactic bulge from
Kent, Dame, \& Fazio (1991) and Freudenreich (1998) for comparison.
As one can see, objects that we observe in
the simulation are very similar to (or perhaps even a little bit larger than)
the bulge of Milky Way, provided, they
convert most of their gas into stars.

Why did we then erroneously conclude in the Introduction that the CDM-type
models predict too low density bulges? The answer to this puzzle is
again illustrated in Fig.\ \ref{figSP}. The horizontal long-dashed line
in that figure shows the central surface density for a homogeneous
top-hat sphere, i.e.\ for a spherical object with the constant density
of 200 times the average density of the universe and with a radius equal
to the virial radius of object A (all four objects have quite similar
virial radii). One can immediately see that the top-hat model underestimates
the central density of an object by about two orders of magnitude! Even
the density at the characteristic radii ($R_C$, $R_e$) is 30--100 times
greater than the fiducial top-hat virial density.

\def\capRS{Evolution of the characteristic radii ({\it lower panel\/})
and densities ({\it upper panel\/}) for the
four objects: 
A ({\it solid lines\/}),
B ({\it long-dashed lines\/}),
C ({\it short-dashed lines\/}), and
D ({\it dotted lines\/}). The horizontal shaded areas mark the range of values
for the Galactic bulge, and the vertical line marks the boundary $z=3.8$
beyond which our simulation fails to resolve cores of the galactic bulges.}
\placefig{
\begin{figure}
\insertfigure{\figdir/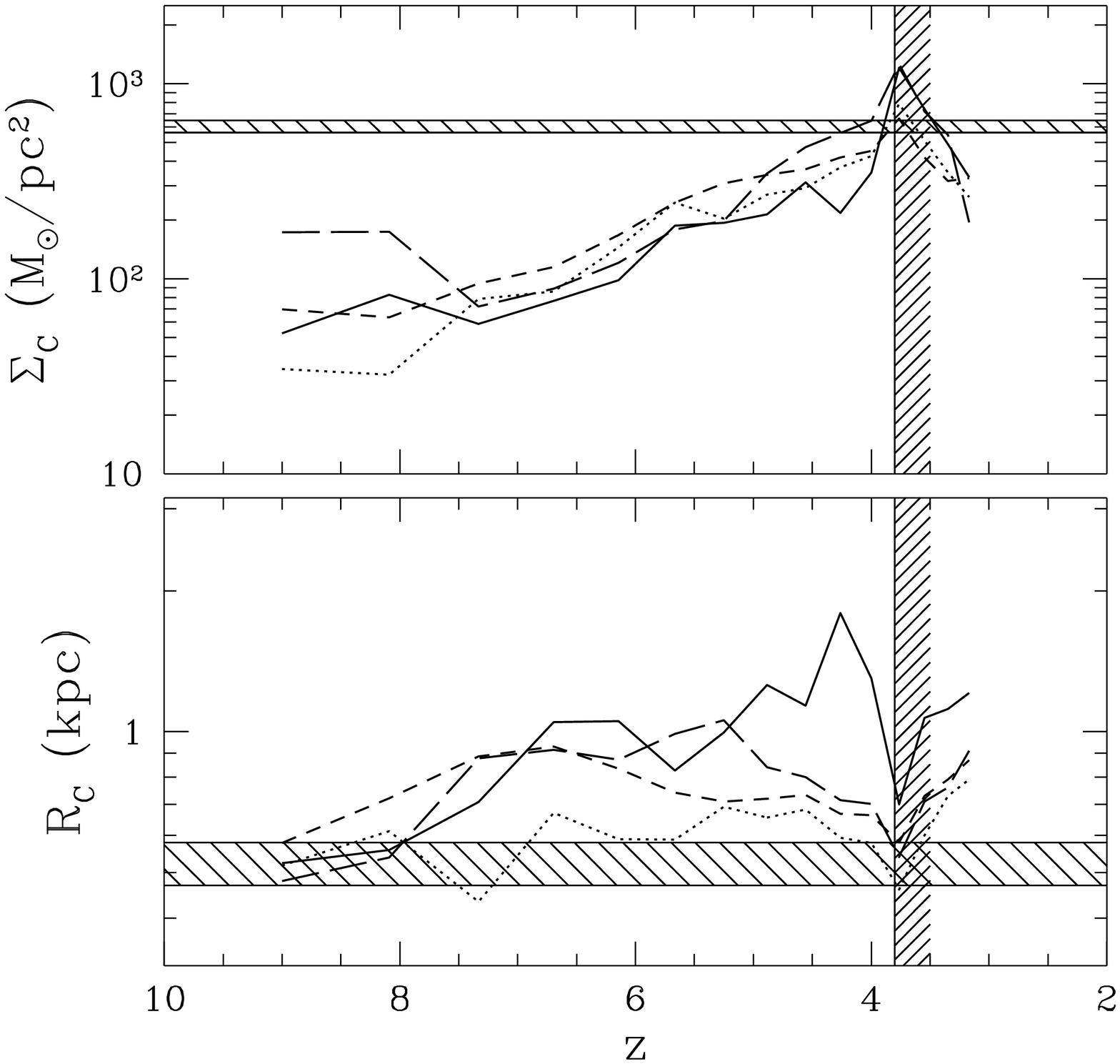}
\caption{\label{figRS}\capRS}
\end{figure}
}
Evolution of the two fit parameters, the characteristic radius and density,
is shown in Fig.\ \ref{figRS} for the four most massive objects in the 
simulation. The densities increase steadily as the gas continues to
accrete and cool inside the dark matter halos, whereas characteristic
radii stay approximately constant for all objects.

\def\capAR{Axis ratios for the dark matter 
({\it solid lines\/}), total baryons (gas and stars, {\it dashed lines\/}), 
and stars ({\it dotted lines\/}) of the
four most massive objects at $z=3.8$ as a 
function of radius in physical (not comoving) units.}
\placefig{
\begin{figure}
\insertfigure{\figdir/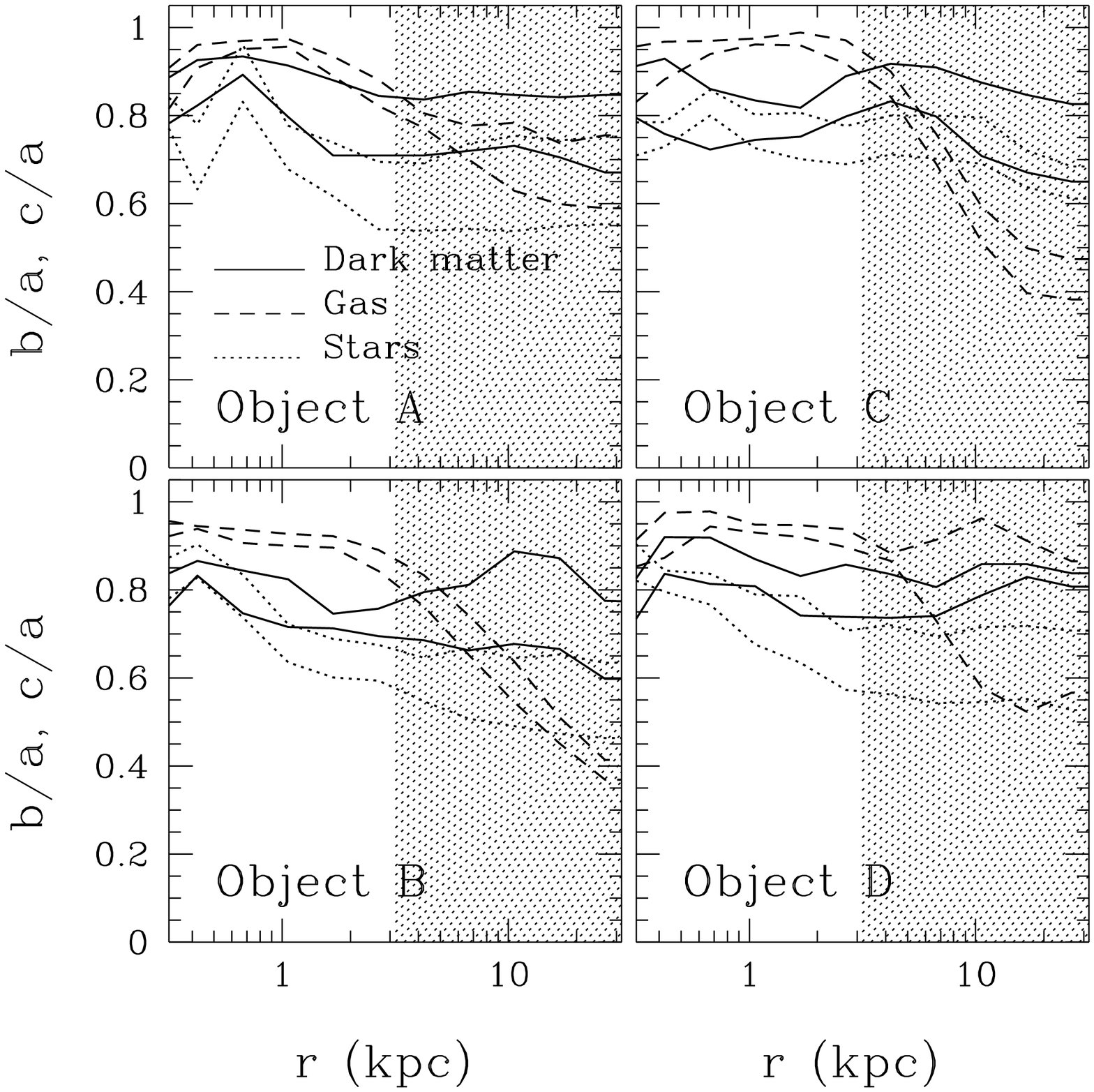}
\caption{\label{figAR}\capAR}
\end{figure}
}
Finally, if we want to demonstrate that we can form galactic bulges in
a realistic cosmological simulation, we should address the question of
the bulge shape. The bulge of our Galaxy is quasi-spherical, or, at 
the very least,
slightly ellipsoidal. Is this shape also reproduced in the simulation?
Figure \ref{figAR} addresses this question. In it we show the
axis ratios for the dark matter, gas, and stars for our four
objects as a function of radius. One can see that in the central parts
the gas and the stellar distributions are quite close to spherical.
Shapes at larger radii, $r>3\dim{kpc}$, shown shaded in Fig.\ \ref{figAR},
vary significantly among all
objects, but at those large distances the gas is far from equilibrium,
and the minute shape of its distribution has little relation to its
final state. The fact that the objects we observe in the simulation are
more-or-less spherical, rather than disk-shaped, is due to the fact that
at high redshift the slope of the linear power spectrum of the density 
fluctuations is close to $-3$. This results in a range of scales becoming
nonlinear almost at the same time, which, in turn, leads to heavy merging
among objects observed in the simulation. Thus, gaseous disks do not have
enough time to form, and the shape of the objects remains quasi-spherical.
Only at later times, when the rate of merging falls down, can a gaseous
disk form inside an object.

\def\capVP{Velocity dispersion and gas temperature profiles for the
four most massive objects at $z=3.8$. Shown are the dark matter
velocity dispersion 
({\it solid lines\/}), gas temperature ({\it long-dashed lines\/})v, gas
velocity dispersion ({\it short-dashed lines\/}), 
and stellar velocity dispersion ({\it dotted lines\/}). The velocity 
dispersions are converted into temperature units.}
\placefig{
\begin{figure}
\insertfigure{\figdir/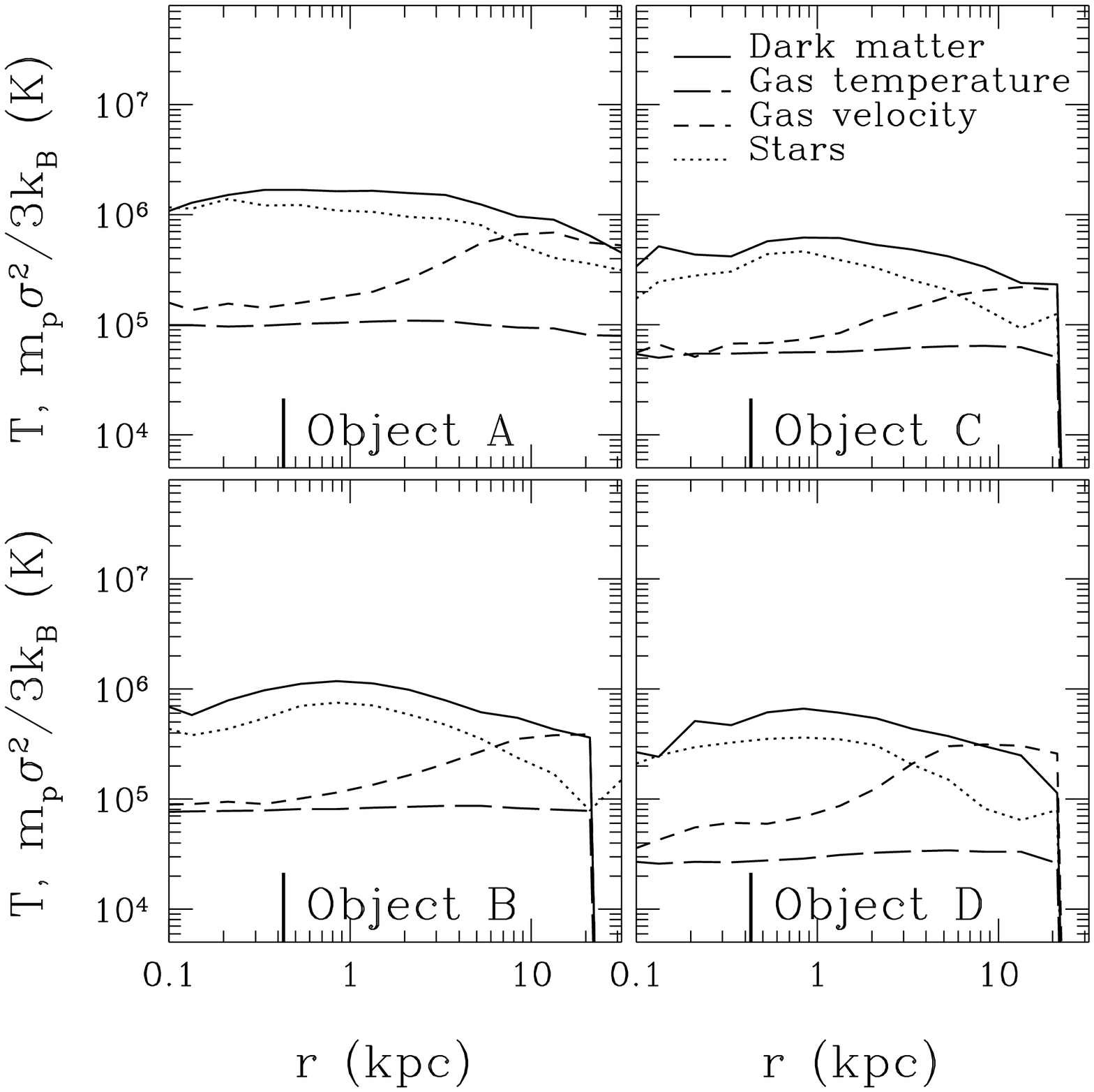}
\caption{\label{figVP}\capVP}
\end{figure}
}
In order to confirm this claim, we show in Figure \ref{figVP} the velocity 
dispersion and temperature profiles for the four objects mentioned above.
As one can see, the gas velocity dispersion is somewhat larger than the
gas temperature, and so the turbulent motions in the gas provide more than
50\% of the support against the gravity. because the molecular and/or
magnetic viscosity is not large on the scales which we can resolve, it would
take some time before the gas can settle into a rotationally supported disk.
Thus, in order for our objects to resemble stellar bulges at the present
epoch, the turbulence dumping time should be large compared to the star
formation time.

\section{Conclusions}

We have showed that objects that form in a realistic cosmological
simulation of a CDM-type cosmological model do look similar to
bulges of normal galaxies. Our objects are still 75\% gaseous
at $z\approx4$, but they form stars at a high rate, and when most of the
gas in those objects will be converted into stars by redshift $z=2$.
In this time frame they resemble the numerous small blue galaxies seen in the
HDF. The number density of such objects predicted from our simulation
is about a factor of two larger than the actual number observed in the
HDF and LBG galaxy samples (Steidel et al.\ 1999), however the precise 
comparison with the observations depends crucially on the (highly
uncertain) assumptions about the star formation, and is not possible at the
this moment.
Later  
the simulated objects will look like
slightly ellipsoidal stellar objects with the density profiles well fit
by the exponential profile and with the parameters of the fit similar to
the parameters of the Milky Way bulge and small elliptical galaxies.

We thus conclude that currently favorable CDM-type cosmological models
have no difficulty in reproducing observed properties of galactic bulges.
On the contrary, models that have galaxy formation at $z\sim30$ 
(Peebles 1997) would form bulges that are two to three orders
of magnitude more dense than the observed ones. This conclusion is further
boosted by the consideration that, due to the limited mass and spatial
resolution of our simulation, we can only {\it underpredict\/} the
densities and masses of cosmological objects.
With several bulges observed in approximately correct mass range (note the
local group with M31, M33, and the Galaxy), and a total luminosity of about
$L_*$ in the volume of $27h^{-3}\dim{Mpc}^3$, we conclude that the
model produces approximately the correct density of bulges.
  
We are grateful to the referee Andreas Burkert for the valuable comments
that significantly improved the paper.
This work was supported in part by the UC Berkeley grant 1-443839-07427
and the NCSA/NSF Subaward 766/ASC97-40300.
Simulations were performed on the NCSA Origin2000 supercomputer under 
the grant AST-970006N.

\placefig{\end{document}}

\clearpage

\tableone

\clearpage

\tableoneb

\clearpage

\tabletwo

\clearpage

\newcounter{figurecap}
\setcounter{figurecap}{0}

\begin{center}
\bf Figure Captions
\end{center}

\refstepcounter{figurecap}
Fig.\ \thefigurecap---\label{figCD}\capCD

\refstepcounter{figurecap}
Fig.\ \thefigurecap---\label{figDP}\capDP

\refstepcounter{figurecap}
Fig.\ \thefigurecap---\label{figMZ}\capMZ

\refstepcounter{figurecap}
Fig.\ \thefigurecap---\label{figSP}\capSP

\refstepcounter{figurecap}
Fig.\ \thefigurecap---\label{figRS}\capRS

\refstepcounter{figurecap}
Fig.\ \thefigurecap---\label{figAR}\capAR

\refstepcounter{figurecap}
Fig.\ \thefigurecap---\label{figVP}\capVP

\end{document}